\documentclass[fleqn,10pt]{wlscirep}
\usepackage[utf8]{inputenc}
\usepackage[T1]{fontenc}

\usepackage{graphicx}
\usepackage{amssymb}
\usepackage{multirow}
\usepackage[caption=false]{subfig} 
\usepackage{pifont}
\usepackage{placeins}
\usepackage{amsmath} 
\usepackage{url}  
\usepackage{cite}
\usepackage{rotating}

\title{Identifying hidden coalitions in the US House of Representatives by optimally partitioning signed networks based on generalized balance}

\author[1,2,*]{Samin Aref}
\author[3]{Zachary P. Neal}
\affil[1]{Max Planck Institute for Demographic Research, 18057 Rostock, Germany}
\affil[2]{Department of Mechanical and Industrial Engineering, University of Toronto, Toronto, ON M5S3G8, Canada}
\affil[3]{Department of Psychology, Michigan State University, East Lansing, MI 48824, USA}
\affil[*]{aref@mie.utoronto.ca}

\begin{abstract}
	In network science, identifying optimal partitions of a signed network into internally cohesive and mutually divisive clusters based on generalized balance theory is computationally challenging. We reformulate and generalize two binary linear programming models that tackle this challenge, demonstrating their practicality by applying them them to partition networks of collaboration in the US House of Representatives. These models guarantee a globally optimal network partition and can be practically applied to signed networks containing up to 30,000 edges. In the US House context, we find that a three-cluster partition is better than a conventional two-cluster partition, where the otherwise hidden third coalition is composed of highly effective legislators who are ideologically aligned with the majority party.
	
	\textbf{Keywords:} Signed network, Polarization, Legislative effectiveness, Generalized balance, Graph partitioning, Optimization
	
	This is a post-peer-review, postprint version of an article published in Scientific Reports. The publisher authenticated version is available online at: \url{http://dx.doi.org/10.1038/s41598-021-98139-w}
\end{abstract}
\begin{document}
	
	\flushbottom
	\maketitle
	\thispagestyle{empty}
	
	\section*{Introduction}
	Signed networks, in which nodes can be connected by positive or negative ties, occur in many contexts. To identify communities in signed networks it is often useful to put the nodes into clusters so that most positive ties are within clusters, while most negative ties are between clusters. Identifying clusters of nodes that optimally meet these criteria is computationally challenging, but we present practical methods for doing so. Applying these new global optimization methods to signed networks of the US House of Representatives shows that legislators are actually organized into three coalitions whose ideological composition offers new insights on the otherwise obscured interplay between partisanship and legislative effectiveness.
	
	Signed networks are studied in a diverse range of contexts in both the natural \cite{iacono_determining_2010,aref2017balance,tahmassebi2019determining} and social \cite{souto2018capturing,neal2018,aref2020multilevel,schoch2020legislators} sciences. Across these contexts, it is often of interest to identify clusters of nodes that are internally cohesive and mutually divisive, and thus partially satisfy the conditions of \textit{generalized balance} \cite{davis1967clustering,cartwright1979balance_and_clusterability,batagelj1994semirings,doreian_partitioning_1996}. Recent computational work on signed network analysis has focused on determining the network's level of balance in general \cite{facchetti2011,sun2014fast,aref2015measuring,aref2017computing}, and in the context of signed graphs with node attributes \cite{he2020energy,du2016structural}. However, although optimization-based methods exist for estimating a network's level of balance \cite{cartwright_structural_1956} by heuristically partitioning it into $k=2$ clusters \cite{sun2014fast} or computing its exact level of balance by optimally partitioning it into $k=2$ clusters \cite{aref2016exact,aref2017balance,aref2020detecting}, identifying an optimal partition of nodes into $k\geq2$ clusters that corresponds to the network's level of \textit{$k$-balance} (a.k.a. \textit{weak balance}, generalized balance, and \textit{clusterability}\cite{davis1967clustering}) has remained a challenge. This computational challenge involves solving fundamental non-deterministic polynomially acceptable hard (NP-hard) graph optimization problems to global optimality\cite{bansal2004correlation,demaine2006correlation,brusco_k-balance_2010,aref2016exact}.
	
	A common misconception about solving NP-hard optimization problems is that they can be addressed using ``only heuristic methods'' \cite{traag_community_2009}. Previous work in this area has used a modified concept of network \textit{modularity} to incorporate signed edges into a modularity maximization procedure \cite{gomez2009analysis,traag_community_2009}. They used a \textit{tabu search} heuristic algorithm on a signed network with $1131$ edges \cite{gomez2009analysis} and used a \textit{simulated annealing} heuristic algorithm on a signed network with $2517$ edges \cite{traag_community_2009}, in each case settling for sub-optimal partitions whose distance from optimality remains unknown. Unlike modularity, the concept of \textit{frustration} \cite{zaslavsky_balanced_1987,aref2017computing} requires no modification for application in signed networks because it originates from Ising models of atomic magnets in which couplings of opposite nature exist \cite{Sherrington} which are analogous to signed ties. Using frustration and two mathematical optimization models, we propose and demonstrate a general method for finding a \textit{globally optimal} partition of signed networks into $k\geq2$ clusters.
	
	Identifying an optimal partition of nodes into internally cohesive and mutually divisive clusters involves two computational challenges. The first challenge is finding a $k$-partition of a signed network, placing nodes into $k$ clusters that minimize intra-cluster negative and inter-cluster positive edges (\textit{frustrated edges}), where $k$ is selected in advance \cite{aref2016exact}. A second challenge is finding the smallest number of clusters $k^*_\text{min}$ that minimizes frustrated edges among all partitions across all values of $k$. These challenges are unique from, but conceptually analogous to related challenges in community detection in unsigned networks: It is difficult to find a modularity maximizing partition into a specific number of clusters, but even harder to find the modularity maximizing partition into any number of clusters \cite{fortunato_community_2010}. We solve the first challenge by generalizing a mathematical programming model for finding an optimal 2-partition \cite{aref2017computing,aref2016exact} and introducing a generalized model to find optimal $k$-partitions. Then, we tackle the second challenge by reformulating another mathematical model \cite{demaine2006correlation} for non-complete graphs and solving it without providing the number of clusters.
	
	We demonstrate the practicality of these methods, and illustrate how they can generate novel insights, by applying them to signed networks of political collaboration and opposition in the US House of Representatives from 1981 to 2018. Research on and descriptions of the US House usually place legislators into clusters defined by legislators' political party affiliations. However, reliance on a simple binary attribute risks oversimplifying this complex system because it ignores information about the positive and negative interactions between individual legislators. We explore whether placing legislators into optimal clusters defined by their interactions, rather than simply by their parties, better captures the coalitional structure of the chamber. We find that the best fitting parsimonious solution places legislators into three clusters characterized by a large liberal coalition, a large conservative coalition, and a smaller ideologically fluid coalition. Interestingly, we find that members of this ideologically fluid third coalition are substantially more effective at passing legislation than members of either dominant coalition. These findings suggest that, although political parties are clearly influential in US politics, some of the heavy lifting in the US House is done by a small splinter coalition of highly effective legislators who are ideologically aligned, but not necessarily collaborating, with members of the majority party's core.
	
	\section*{Partitioning signed networks}
	In this section, after introducing notions of $k$-balance and signed networks, we propose two related mathematical models. The first model finds an optimal partition of nodes into exactly $k$ clusters. The second model finds an optimal partition across all possible partitions. When used together, these models \textit{guarantee a globally optimal partition} and provide the smallest number of clusters according to generalized balance.
	
	\subsection*{Preliminaries}
	\label{preliminaries}
	A \textit{signed network} is an undirected simple graph with positive and negative signs on the edges usually denoted as $G = (V,E,\sigma)$ where $V$ and $E$ are the sets of nodes and edges respectively, and $\sigma$ is the sign function $\sigma: E\rightarrow\{-1,+1\}$. Graph $G$ contains $|V|=n$ nodes and its symmetric signed adjacency matrix is denoted by $\textbf{A}$. The set $E$ of edges contains $m^-$ negative edges and $m^+$ positive edges adding up to a total of $|E|=m=m^+ + m^-$ undirected signed edges. An edge with endpoints $i$ and $j$ is represented by $(i,j)$ such that $i<j$. Given a signed graph $G=(V,E,\sigma)$, a $k$\textit{-partition} is a division of the nodes $V$ into $k$ non-empty subsets $V_1,V_2,\dots,V_k$ such that $V_i \cap V_j=\varnothing \forall i \neq j$ and $\cup_{i=1}^k V_i=V$ (i.e.\ every node belongs to exactly one subset). 
	
	Balance theory was conceptualized in the 1940s in the context of social psychology \cite{heider_social_1944}, recast in graph theoretic terms in the 1950s \cite{cartwright_structural_1956}, and generalized in the 1960s \cite{davis1967clustering}. Whereas classic balance holds that a signed network can be partitioned into up to two clusters \cite{cartwright_structural_1956}, generalized balance holds that it can be partitioned into any number of clusters. Generalized balance theory allows a more flexible structural decomposition of networked systems, which in turn offers a more nuanced view of polarization in social and political systems \cite{layman2006,zhang2008,moody2013}. According to generalized balance theory, a signed network is \textit{$k$-balanced} (i.e.\ \textit{clusterable}) if its nodes can be partitioned into $k$ clusters (or ``coalitions'' \cite{harary_simple_1980}) such that each positive edge joins nodes belonging to the same cluster, and each negative edge joins nodes belonging to different clusters \cite{davis1967clustering}. Edges that fail to meet these criteria (i.e. a negative edge within a cluster, or positive edge between clusters) are called \textit{frustrated} edges under that partition.
	
	Generalized balance in empirical signed networks can be analyzed by measuring their distance to clusterability \cite{cartwright1979balance_and_clusterability,doreian_partitioning_1996,aref2016exact}. The distance of a given network $G$ to clusterability can be quantified as the minimum number of frustrated edges among all possible partitions into $k$ clusters \cite[$k$-\textit{clusterability index} $C_k(G)$]{doreian_partitioning_1996}, or the minimum number of frustrated edges among all possible partitions with any number of clusters $1\leq k \leq n$ \cite[\textit{clusterability index} $C(G)$]{cartwright1979balance_and_clusterability}. Obtaining these measures require intensive computation and are NP-hard \cite{bansal2004correlation}.
	
	\begin{figure}[ht]
		\centering
		\includegraphics[width=12cm]{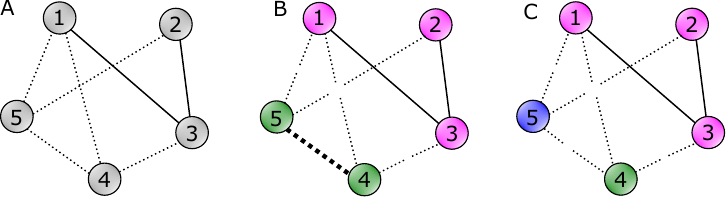}
		\caption{(A) An example signed network. (B) Evaluating classic balance via bi-partitioning involves finding the optimal 2-partition $\{\{1,2,3\},\{4,5\}\}$ and the edge $(4,5)$ which is frustrated under it. (C) Evaluating generalized balance and clusterability via $k$-partitioning leads to the optimal 3-partition $\{\{1,2,3\},\{4\},\{5\}\}$ which reduces the frustrated edges to $0$.}
		\label{fig:toy}
	\end{figure}
	
	Fig.\ \ref{fig:toy}(A) shows an example signed network with five negative edges (dotted lines) and two positive edges (solid lines). The signed network can be optimally partitioned into two clusters based on classic balance (B), or three clusters based on generalized balance (C). The classic approach leads to the 2-partition $\{\{1,2,3\},\{4,5\}\}$ (shown by green and purple colors in Fig.\ \ref{fig:toy}B) which minimizes the total number of intra-cluster negative and inter-cluster positive edges to $C_2(G)=1$. The generalized approach, (Fig.\ \ref{fig:toy}C), leads to the 3-partition $\{\{1,2,3\},\{4\},\{5\}\}$ which satisfies the conditions of generalized balance ($C(G)=C_3(G)=0$). 
	
	\subsection*{Finding an optimal $k$-partition and the $k$-clusterability index}
	\label{subsec:k-cluster}
	We formulate an optimization model that computes the $k$-clusterability index of an input signed network in its optimal objective function. In a given feasible solution of the optimization problem, each node belongs to one of a set of $k$ clusters $C=\{1, 2, \dots, k\}$. The binary decision variable $x_{ic}$ takes the value $1$ if node $i \in V$ belongs to cluster $c\in C$ (and $x_{ic}=0$ otherwise).
	
	We consider that a positive edge $(i,j) \in E^+$ is frustrated (indicated by $f_{ij}=1$) if its endpoints $i$ and $j$ are in different clusters%, i.e.,\ $c_i \ne c_j$
	; otherwise it is not frustrated (indicated by $f_{ij}=0$). A negative edge $(i,j) \in E^-$ is frustrated (indicated by $f_{ij}=1$) if its endpoints $i$ and $j$ are in the same cluster%$c_i = c_j$
	; otherwise it is not frustrated (indicated by $f_{ij}=0$). 
	
	Using the binary decision variable $x_{ic}$, we formulate the process to find an optimal $k$-partition and compute the $k$-clusterability index as the binary linear programming model in Eq.\ \eqref{eq1}. The model in Eq.\ \eqref{eq1} is an extension of a model based on classic balance which provides an optimal 2-partition and computes the 2-clusterability index (a.k.a.\ the frustration index) of a signed network \cite{aref2017computing,aref2016exact}.
	
	\begin{equation} \label{eq1}
		\begin{split}
			\min \sum_{(i,j) \in E}&   f_{ij}  \\
			\text{s.t.} \quad \sum_{c \in C} x_{ic} &= 1 \quad \forall i \in V \\
			f_{ij}  &\ge  x_{ic} - x_{jc} \quad \forall (i,j) \in E^+,  ~\forall c \in C \\
			f_{ij}  &\ge  x_{ic} + x_{jc} -1 \quad \forall (i,j) \in E^-,  ~\forall c \in C \\
			x_{ic} &\in \{0,1\} \quad  \forall i \in V,  ~\forall c \in C \\
			f_{ij} &\in \{0,1\} \quad \forall (i,j) \in E 
		\end{split}
	\end{equation}
	
	The objective function in Eq.\ \eqref{eq1} computes the minimum number of frustrated edges among all $k$-partitions. The first set of constraints in Eq.\ \eqref{eq1} ensures that each node belongs precisely to one cluster. The second and third sets of constraints formulate the relationship between frustration of an edge (left-hand side) and the cluster membership of the endpoints of that edge (right-hand side) respectively for positive edges and negative edges. Refer to the Supplementary Information for more details and an illustrative numerical example on how the $k$-partitioning model in Eq.\ \eqref{eq1} works.
	
	\subsection*{Finding an optimal partition and the clusterability index}
	\label{subsec:cluster}
	The more general problem of finding an optimal partition without specifying $k$ and computing the clusterability index of a signed network $G$ is known as the Correlation Clustering problem \cite{bansal2004correlation} (and the Clique Partitioning problem if the graph is complete \cite{mehrotra1998cliques}). We reformulate the mathematical model initially proposed in \cite{demaine2006correlation} which is defined in the context of complete graphs and widely used in the literature \cite{figueiredo2013mixed,drummond2013efficient,levorato2015ils,levorato2017evaluating} as follows. For every pair of nodes $i,j,i<j$, we define the binary decision variable $y_{ij}$ which takes the value 1 if $i$ and $j$ belong to the same cluster and takes the value 0 otherwise. 
	
	\begin{equation} \label{eq2}
		\begin{split}
			\min \sum_{(i,j) \in E}&  a_{ij}((a_{ij}+1)/2) - a_{ij}y_{ij}  \\
			\text{s.t.} \quad 
			y_{ij} + y_{ik}   &\ge  1 + y_{jk} \quad \forall (i,j,k) \in T \\
			y_{ij} + y_{jk}   &\ge  1 + y_{ik} \quad \forall (i,j,k) \in T \\
			y_{ik} + y_{jk}   &\ge  1 + y_{ij} \quad \forall (i,j,k) \in T \\
			y_{ij} &\in \{0,1\} \quad  \forall i \in V,  ~j \in V ,~i<j \\
		\end{split}
	\end{equation}
	
	The model in Eq.\ \eqref{eq2} uses these binary variables to count the frustrated edges in the objective function. in Eq.\ \eqref{eq2}, the term $a_{ij}$ represents the entry of the input graph's adjacency matrix $\textbf{A}$ associated with the pair of nodes $i,j \in V$. To efficiently handle possibly non-complete graphs, we use the set $T$ for the constraints of the model in Eq.\ \eqref{eq2}. $T=\{(i,j,k)\in V^3 \mid  |a{_i}{_j}|+|a{_i}{_k}|+|a{_j}{_k}| \ge 1, i<j<k \}$ denotes the set of all node triples with at least one edge between two of them. Refer to the Supplementary Information for more details and an illustrative numerical example on how the partitioning model in Eq.\ \eqref{eq2} works.
	
	Although we use both models in Eq.\ \eqref{eq1}--\eqref{eq2}, they are not necessarily dependent. Under the assumption that $k<<<n$, our proposed model in Eq.\ \eqref{eq1} is less computationally intensive than the model proposed by \cite{demaine2006correlation}, which we have reformulated in Eq.\ \eqref{eq2}. Despite similar scaling of the number of variables with $\mathcal{O}(n^2)$, constraints of \eqref{eq1} have a quadratic growth with $\mathcal{O}(n^2)$ while constraints of \eqref{eq2} have a cubic growth $\mathcal{O}(n^3)$.
	
	These models can be used for optimally partitioning any signed network into internally cohesive and mutually divisive clusters based on generalized balance. However, it is important to note that they can yield a multiplicity of optimal solutions, that is, they do not necessarily yield a single unique partition because multiple optimal solutions may exist (see the \textit{Supplementary Information} for more details). Despite this potential multiplicity, these models offer two kinds of advantages over existing methods with similar goals. First, unlike heuristic partitioning methods that can provide locally optimal partitions \cite{traag_community_2009}, the partitions identified by these models come with a \textit{guarantee} of \textit{global optimality} that means no better partition exists. Second, unlike other optimal partitioning methods that have been applied to small \cite{brusco_k-balance_2010,figueiredo2013mixed} or complete \cite{bansal2004correlation} signed networks, these models can be practically solved even for networks of considerable size and order, and for networks that are not complete, which are typical in social contexts. In the next section we demonstrate their practicality using networks with up to $30,000$ edges. We solve the optimization models in Eq.\ \eqref{eq1} and \eqref{eq2} to global optimality using \textit{Gurobi} solver (version 9.1) \cite{gurobi} on a virtual machine with 32 Intel Xeon CPU E7-8890 v3 @ 2.50 GHz processors running 64-bit Microsoft Windows Server 2019 R2 Standard. 
	
	\section*{Partitioning the US House networks}
	In the previous section we generalized one model and reformulated another model that together guarantee a globally optimal partition of a signed network according to generalized balance. In this section, we show that they are computationally feasible and can be solved in a practical amount of time. To illustrate their practicality, we apply them to 19 networks varying in size, density, and structure that represent political collaborations in the US House of Representatives in different eras. Although these networks are not `large' compared to some networks ($n \sim 445$, $4954 \leq m \leq 31936$), they are large by comparison to the size of signed networks for which globally optimal partitions have been obtained before \cite{brusco_k-balance_2010,figueiredo2013mixed,arinik2021multiplicity}.
	
	\subsection*{Optimal coalitions}
	\label{subsec:result1}
	We compare several ways to partition US House legislators into clusters or ``coalitions'' \cite{harary_simple_1980}, with the goal of determining the optimal number and the composition of these coalitions. The fitness of a given partition is indicated by its associated number of frustrated edges. The conventional method is to partition legislators into coalitions based on their party affiliations, while here we also explore partitioning legislators into coalitions by applying the optimization models in Eq.\ \eqref{eq1}--\eqref{eq2} to signed networks of their collaborations and oppositions. Throughout our application of these models in the US House context, we use the term ``coalition'' to refer to the clusters of legislators within a partition, however the partition is obtained, not only because it is commonly used in political contexts, but also because it was the term suggested for signed network partitions by Harary and Kabell~\cite{harary_simple_1980}. Legislators' memberships in these coalitions depend on either an attribute (e.g. their political party affiliation) or the solution to \eqref{eq1}--\eqref{eq2}, but does not necessarily imply their cohesion with other members of the same coalition.
	
\begin{figure}
		\centering
		\includegraphics[width=12cm]{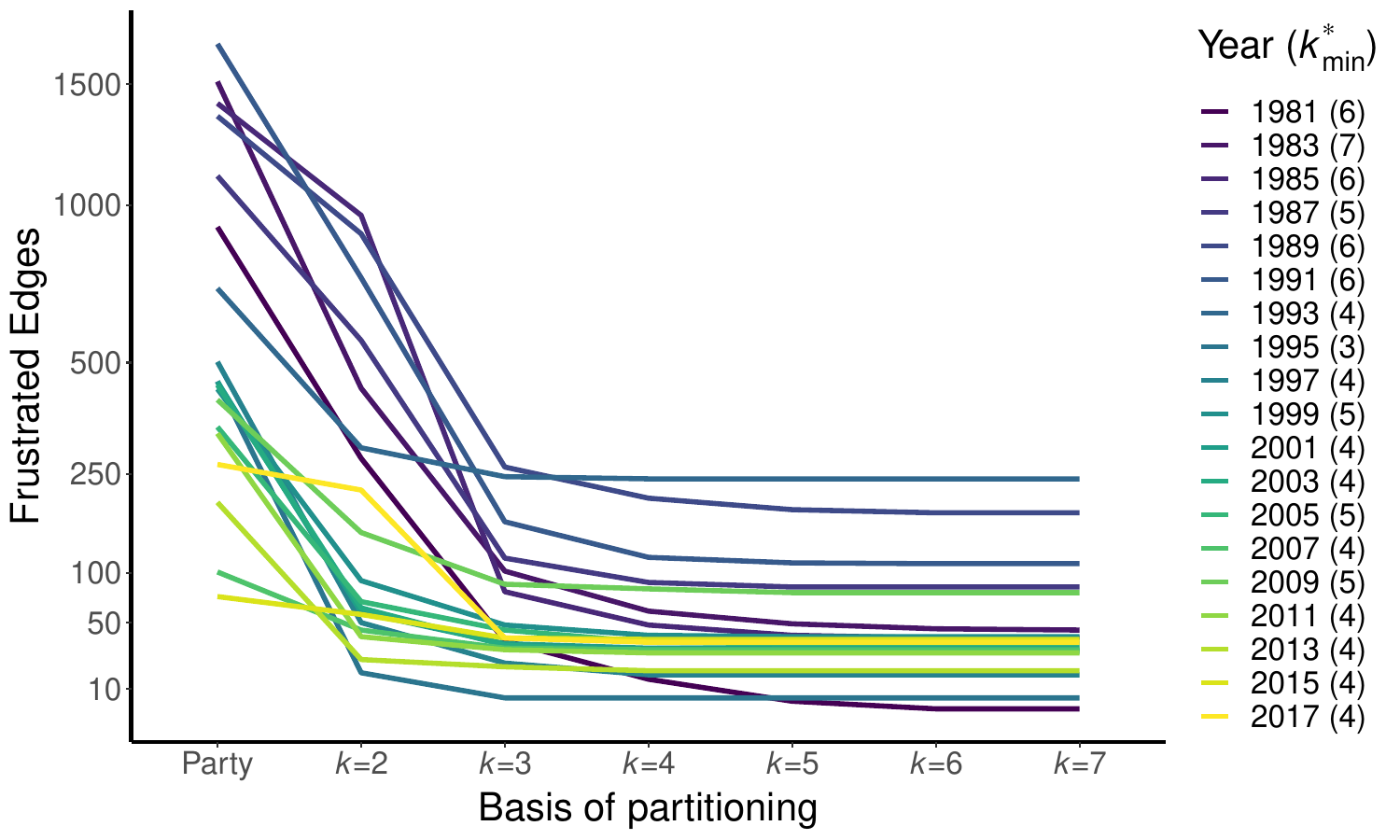}
		\caption{Number of frustrated edges (y-axis) of signed US House networks partitioned using different criteria (x-axis). Each line represents a single network, corresponding to a session of the US House starting in the given year. Fewer frustrated edges indicate that the partition is more consistent with the ties of collaboration and opposition between legislators.}
		\label{fig:k-clusterability}
	\end{figure}

	Fig. \ref{fig:k-clusterability} illustrates the number of frustrated edges (y-axis) for partitions based on party affiliations and optimal $k$-partitions for $k \in \{2,3,\dots,7\}$ (x-axis) in signed US House networks (see \textit{SI} Table 1). The number of frustrated edges for a party-based partition (denoted by $C_{\text{party}}(G)$) is considerably larger than that of an optimal 2-partition. This implies that defining coalitions simply in terms of legislators' party affiliations leads to many frustrated edges, and therefore to a poor description of the coalition structure of the chamber. The number of frustrated edges decreases further from $k=2$ to $k=3$, which implies that defining coalitions in terms of classic balance still leads to many frustrated edges and thus a poorer fit than defining coalitions in terms of generalized balance. For $k>3$ there is only marginal decline, and then stagnation, in the number of frustrated edges. Substantively, these results suggest that the signed US House networks are better described by a partition into $k>2$ coalitions than by a more conventional partition into only two coalitions \cite{aref2020detecting}.
	
	Fig. \ref{fig:k-clusterability} also reveals the changes over different eras of the House (e.g. sessions with start years 1981-1993 in darker blue-purple shades and 2003-2017 sessions in lighter green-yellow shades). Party-based partitions offer a better fit (i.e. fewer frustrated edges) in recent sessions than in earlier sessions due to increases in partisanship \cite{neal2018,andris_2015,aref2020detecting}. However, despite changes in the level of partisanship over time, for every session $C_{\text{party}}(G) > C_2(G) > C_3(G)$.
	
	Because the results from Fig. \ref{fig:k-clusterability} only cover a small range of $k$, a natural question is whether the fit could be improved further by using larger values of $k$. Finding the answer is not practically feasible using only the model in Eq.\ \eqref{eq1}. Therefore, we solve the model in Eq.\ \eqref{eq2} to find the minimum number of frustrated edges, $C(G)$, across all possible partitions for all possible values of $1\leq k \leq n$. By juxtaposing $C_k(G)$ from \eqref{eq1} and the values of $C(G)$ from \eqref{eq2}, we determine whether the low-points observed in Fig. \ref{fig:k-clusterability} represent the clusterability indices $C(G)$ of the networks or the number of frustrated edges could decline further for $k>7$. 
	
	Through this comparison, we verify that further decline in the frustrated edges is not possible because among all 19 networks, $C(G) = C_k(G)$ at $k\leq7$. The legend of Fig. \ref{fig:k-clusterability} shows for each network the exact \textit{point of stagnation} $k^*_{\text{min}}$, which is the smallest number of clusters that minimizes the $k$-clusterability index across all values of $k$: $k^*_{\text{min}}=\text{arg} \min_{1 \leq i \leq n} C_i(G)$.
	
	\subsection*{Coalition ideology}
	\label{subsec:ideology}
\begin{figure}[ht]
		\centering
		\includegraphics[width=10cm]{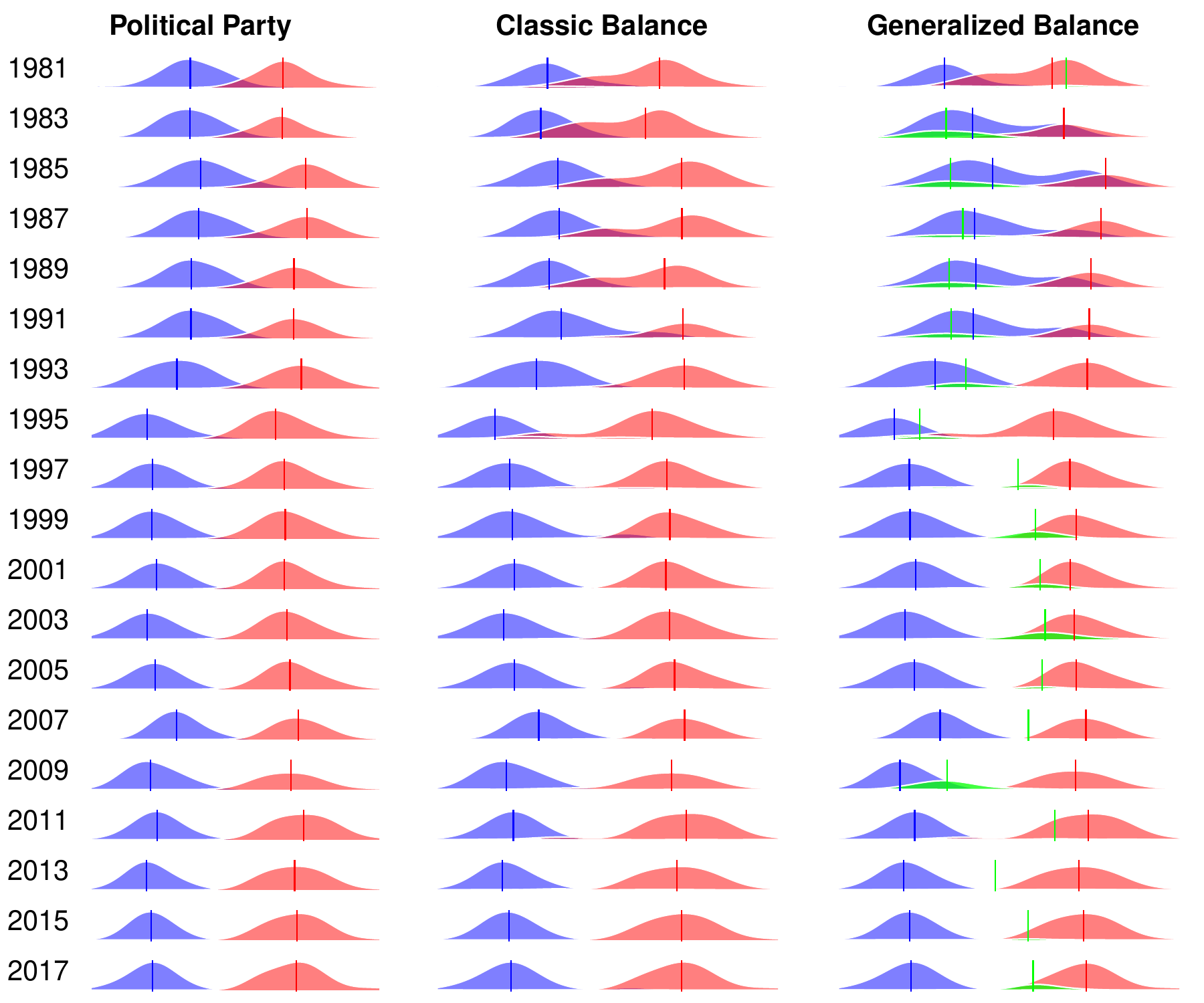}
		\caption{Distribution of coalition members' ideology in the US House of Representatives. Blue (red) curves indicate the ideologies of Democrats (Republicans) in the left column and that of the dominant liberal (conservative) coalitions in the center and right columns. In the right column, green curves indicate the ideologies of members of the smallest coalition.}
		\label{fig:ridge}
	\end{figure}

	Having identified several ways to assign legislators to coalitions in the US House, including optimal $k$-partitions and optimal partitions, we now examine the ideological compositions of coalitions defined from three perspectives: party, classic balance ($k=2$), and generalized balance ($k=3$). Although we found that $3 \leq k^*_{\text{min}} \leq 7$, in the remaining substantive analyses we focus on the 3-partition in the generalized balance case because $k>3$ offers only small improvements in fit and therefore $k=3$ offers a reasonable trade-off between fit and parsimony (See \textit{SI} Figures 5 \& 6). Fig. \ref{fig:ridge} displays the distribution of coalition members' ideology, for each method of defining coalitions (See \textit{SI} Table 2). Coalitions with left-leaning liberal ideologies are shaded blue, while coalitions with right-leaning conservative ideologies are shaded in red; the solid vertical lines indicate a coalition's median ideology. 
	
	Partitioning legislators into coalitions based on their political party affiliations (Fig. \ref{fig:ridge}, left column) is the conventional approach in political science, and here displays the familiar pattern of increasing ideological polarization. Partitioning legislators based on classic balance (Fig. \ref{fig:ridge}, center column) offer a more data-driven classification because legislators' coalition memberships are based on their collaborative and oppositional interactions, but is still restrictive because it allows a maximum of two coalitions. The classic balance coalitions display similar ideological distributions to those based on political party: increasing liberal-conservative ideological polarization.
	
	Partitioning legislators into 3 coalitions based on generalized balance (Fig. \ref{fig:ridge}, right column) also offers a data-driven classification, but allows more nuance. Like the other partitions, the generalized balance partition is characterized by a large liberal coalition and and a large conservative coalition that diverge over time. However, it also includes a smaller and ideologically fluid coalition shaded in green. In the 435-member chamber, this `third coalition' ranges in size from only 4 members in the 113\textsuperscript{th} session (2013) to 69 members in the 111\textsuperscript{th} session (2009). It also ranges in ideology from very liberal in the 98\textsuperscript{th}--102\textsuperscript{nd} sessions (1983--1991), to center-left in the 103\textsuperscript{th} and 111\textsuperscript{th} sessions (1993 and 2009), to center-right in the 105\textsuperscript{th}--110\textsuperscript{th} sessions (1997--2007).
	
	\subsection*{Coalition effectiveness}
	\label{subsec:effectiveness}
\begin{figure}[ht]
		\centering
		\includegraphics[width=12cm]{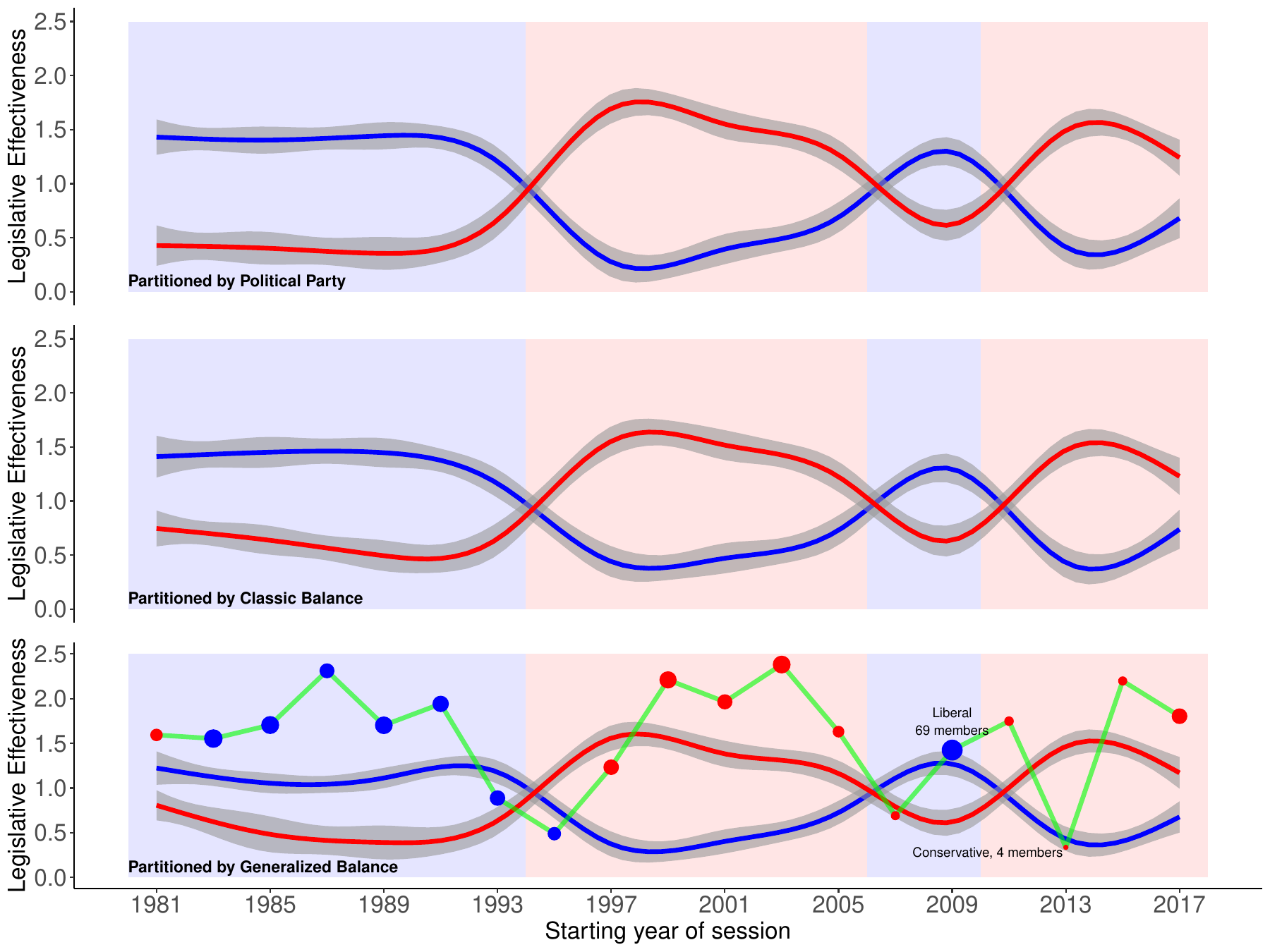}
		\caption{Mean of coalition members' legislative effectiveness in the US House of Representatives. Blue (red) lines indicate the mean legislative effectiveness of Democrats (Republicans) in the top panel and that of the dominant liberal (conservative) coalitions in the center and bottom panels. In the bottom panel, the green line indicates the mean ideological effectiveness of members of the smallest coalition, while the size and color of the dot indicates the size and mean ideology of this coalition. Background shading indicates whether Democrats (blue) or Republicans (red) held a majority in the chamber during the respective session.}
		\label{fig:effectiveness}
	\end{figure}

	The primary task of legislators is to pass laws, and their ability to do so is referred to as \textit{legislative effectiveness} \cite{olson_measures_1972,frantzich_who_1979,volden2014}. Therefore, we examine the legislative effectiveness of coalitions in the US House of Representatives, again considering coalitions defined from three perspectives: party, classic balance ($k=2$), and generalized balance ($k=3$). Fig. \ref{fig:effectiveness} displays coalition members' mean effectiveness, for each method of defining coalitions (See \textit{SI} Table 2). The left-leaning liberal coalition shown as a blue line and the right-leaning conservative coalition shown as a red line. Gray bands illustrate the 95\% confidence interval around each estimate, while the blue (Democrat) and red (Republican) backgrounds indicate the majority party in a given session.
	
	Coalitions based on political parties (Fig. \ref{fig:effectiveness}, top panel) illustrate an expected pattern \cite{moore_legislative_1969}: the majority party is most effective. This occurs not only because the majority party has more votes, but because it controls key procedural details of the chamber including deciding which bills will come for a vote and when (i.e. agenda-setting power \cite{volden2014}). Coalitions based on classic balance (Fig. \ref{fig:effectiveness}, center panel) display essentially the same pattern.
	
	Coalitions based on generalized balance (Fig. \ref{fig:effectiveness}, bottom panel) also display a similar pattern, but with important differences. The large liberal coalition is still more effective when Democrats hold the majority, while the large conservative coalition is still more effective when Republicans hold the majority. However, these two dominant coalitions are both less effective than their party- or classic balance-defined counterparts. These lower levels of effectiveness are explained by the inclusion of the third coalition, shown as a green line, which is the most effective coalition in most sessions. The size and color of the dots along this green line indicate the third coalition's size and median ideology, and highlight that members of the third coalition usually are ideologically aligned with the majority party.
	
	During transitional periods when the majority party changed, members of the third coalition are temporarily less effective. However, during periods of stable party control \cite{mayhew2005}, the highly effective third coalition has been anchored by a small number of consistent and ultra-effective members. For example, the liberal-leaning third coalition during the Democratic-controlled 99\textsuperscript{th}--102\textsuperscript{nd} sessions (1985--1990) was anchored by Rep. Pat Williams (D-MT1, mean effectiveness = 4.49), Rep. Barney Frank (D-MA4, 4.02), and Rep. Daniel Glickman (D-KS4, 3.68). Similarly, the conservative-leaning third coalition during the Republican-controlled 106\textsuperscript{th}--108\textsuperscript{th} sessions (1999--2004) was anchored by Rep. Christopher Smith (R-NJ4, 8.44), Rep. Bill Young (R-FL10, 4.41), and Rep. Nancy Johnson (R-CT6, 2.98).  Most recently, the conservative-leaning third coalition during the Republican-controlled 114\textsuperscript{th}--115\textsuperscript{th} sessions (2015--2018) was anchored by Rep. Edward Royce (R-CA39, 5.46), Rep. John Katko (R-NY24, 5.36), and Rep. Dave Reichert (R-WA8, 2.30).
	
	Not only are members of the third coalition more effective than their traditional liberal and conservative coalition counterparts, but they also maintain distinctive political relations. Members of the traditional coalitions have 2.68 negative edges for every positive edge, but members of the third coalition have 21.18 negative edges for every positive edge (See \textit{SI} Figure S3). Moreover, although 8.44\% of traditional coalition members' negative edges are with co-partisans, over one-quarter (25.6\%) of third coalition members' negative edges are with co-partisans.
	
	\section*{Discussion}
	\label{sec:discussion}
	Optimally partitioning signed networks according to generalized balance theory is computationally challenging, but often essential to understanding their structure. In this paper, we have developed a solution to this challenge, both demonstrating its computational feasibility and highlighting the novel structural insights that the resulting optimal partitions can reveal. Specifically, we have developed a pair of optimization models that make it practical to partition a signed network into exactly $k$ clusters that minimize the number of frustrated edges across all possible $k$-partitions (taking 3.3 hours on average for our networks with up to $\sim 30,000$ edges using Eq.\ \eqref{eq1}), and to identify the smallest number of clusters that minimizes the number of frustrated edges across all possible partitions (taking 14 hours\footnote{taking 13.74 minutes when solving these instances again using Gurobi 9.5.2 on a laptop with 11th Gen Intel(R) Core(TM) i7-11800H @ 2.30GHz and 64 GBs of RAM running Windows 10 Home} on average for our networks with up to $\sim 30,000$ edges using Eq.\ \eqref{eq2}). Applying these models to signed networks of collaboration and opposition among legislators in the US House allowed us to determine that these relationships are not structured by legislators' political party affiliations, but instead by a three coalition system composed of a dominant liberal coalition, a dominant conservative coalition, and a previously obscured `third coalition.' This hidden third coalition is noteworthy because its median ideology is unstable, however its members are consistently more effective at passing legislation than their colleagues in either of the dominant coalitions.
	
	Just as community detection algorithms advanced the ability to uncover patterns in unsigned networks a decade ago \cite{fortunato_community_2010}, these models can advance the ability to uncover patterns in signed networks. However, unlike most community detection algorithms for which global optimization is not possible \cite{lancichinetti2011limits}, our models guarantee an optimal signed network partition. These innovations are important because signed networks are already studied in a wide range of contexts including biology \cite{iacono_determining_2010,aref2017balance,tahmassebi2019determining}, finance \cite{aref2017balance,souto2018capturing}, and politics \cite{neal2018,aref2020detecting,schoch2020legislators}. Moreover, statistical models now exist that enable signed networks to be constructed from virtually any empirical bipartite network data \cite{domagalski2021}, making signed networks available for analysis in a still broader range of contexts. The models we propose are perfectly general, but we demonstrated their practicality for globally optimal partitioning of real-world signed networks with up to $30,000$ edges. In practice, this is a minor limitation because most empirical signed networks contain fewer edges, and models for constructing signed networks include methods for sparsifying otherwise dense signed networks \cite{domagalski2021}.
	
	In addition to the methodological advances that our optimization models offer in the study of signed networks, our illustrative application of these models has also revealed a new way of thinking about how the US House of Representatives is organized. We observe that partitioning legislators into three coalitions according to generalized balance offers a better fit to their observed pattern of collaborations and oppositions than simply clustering them by political party. This suggests that the forces guiding coalition formation in the US House are more subtle and go beyond partisanship alone, even during periods of extreme polarization.
	
	The previously obscured `third coalition' we identified is unique in two important respects. First, members of the third coalition are highly effective at passing legislation, which has implications for how a party's majority status is interpreted. Although members of the majority political party always appear to be more effective than members of the minority party, a substantial portion of this apparent majority advantage is conferred by the highly effective members of the third coalition, who tend to be ideologically aligned with the majority. Second, members of the third coalition have a much higher ratio of oppositions (negative edges) to collaborations (positive edges), and maintain more oppositions with members of their own party, which has implications for how membership in the third coalition is interpreted. These patterns suggest that although members of the third coalition may be ideologically aligned with the dominant coalition and majority party, they nonetheless represent a breakaway faction that are highly effective despite their rejection of partisanship. Our ability to identify such a cluster is noteworthy because it provides empirical support for earlier simulation studies suggesting that the introduction of independent legislators to an existing two-party legislature can increase the body's overall legislative effectiveness \cite{pluchino2011accidental}. Although these simulation studies might have been viewed as hinting at a strategy for reinvigorating democratic systems plagued by partisanship, our findings suggest it may already be in place in the US House of Representatives.
	
	\section*{Methods}
	We infer the collaboration and opposition patterns of legislators from their bill co-sponsorships \cite{fowler_legislative_2006,neal_backbone_2014,neal2018}. These data begin as a bipartite network \textbf{B} in which legislators are connected to the bills they sponsor in a given session. From this, we construct the bipartite projection \textbf{P}, which captures the number of bills each pair of legislators has co-sponsored together. Finally, we use the Stochastic Degree Sequence Model (SDSM) \cite{neal_backbone_2014}, implemented in the \texttt{backbone} package (version 1.5.0) in R \cite{backbone,domagalski2021}, to statistically infer a signed network of political collaboration and opposition. The SDSM applies a statistical test to the bipartite projection to yield a signed backbone \textbf{P$'$} in which there exists a positive (negative) edge between each pair of legislators who have co-sponsored statistically significantly more (fewer) bills than expected by chance. The random expectation is obtained from a canonical null model in which bill sponsorship is random, but expected values of both degree sequences of \textbf{B} are preserved. Because the SDSM involves performing a statistical test for each pair of legislators, we ensure a family-wise error rate of $\alpha=0.01$ by applying a Holm-Bonferroni correction \cite{holm1979simple}.\\
	
	We measure legislators' ideology using 1\textsuperscript{st} dimension Nokken-Poole ideology scores obtained from the Voteview database \cite{lewis2018voteview}. These scores are similar to the widely used DW-Nominate ideological scores \cite{poole1984polarization,poole2000congress,cox2002measuring}, ranging from $-1$ (liberal) to $1$ (conservative), except that they can vary across sessions. We measure legislators' effectiveness using legislative effectiveness scores provided by the Center for Effective Lawmaking at \url{https://thelawmakers.org/data-download}. These scores were computed from fifteen indicators constructed from the intersection of three types of bills (commemorative, substantive, or substantive and significant) and five stages of a bill’s progression through the legislative life cycle (sponsored, committee action, post-committee action, chamber passage, and becoming law). These fifteen indicators capture the effectiveness of a legislator to advance their agenda items using methods described by \cite{volden2014}, and are normalized so that the mean effectiveness in each session is 1.

	\section*{Acknowledgements}
	The authors are thankful to Aliakbar Akbaritabar, Sarah C. Johnson, Oul Han, and David Schoch for comments and discussions, which helped to improve this article.
	
	\section*{Author contributions statement}
	S.A. formulated mathematical models, used new methods to run experiments and solve models, combined and reconfigured data, obtained the results, and assisted in plotting and analyzing them; Z.P.N. developed new methods to infer network data, prepared plots, visualized the networks, and analyzed the results; both authors contributed to reviewing the literature, designing and conducting the research, and writing the paper.
	
	\subsection*{Supplementary information}
	All the data and codes used in this study are publicly available with links and descriptions provided in the supplementary information.
	
	\clearpage
	
	{\centering
		{\LARGE Supplementary information for \\ \textbf{identifying hidden coalitions in the\\ US House of Representatives by optimally partitioning\\ signed networks based on generalized balance} \par}}

	\noindent
	\textbf{\\This PDF file includes:}
	
	\noindent
	Supplementary Text\\
	Figs. 5 to 7\\
	Tables 1 to 2\\
	Caption for Movie S1\\
	Captions for Databases S1 to S2\\
	All References\\
	\\
	\textbf{Other Supplementary Materials for this manuscript include the following:}
	
	\noindent
	Movie S1\\
	Databases S1 to S2
	
	\clearpage

	\section*{Supplementary Text}
	
	\subsection*{Data and code availability}
	All network data, numerical results, and replication code related to this study are publicly available with links provided in this document. %The Python code for the optimization models used is this study will be made publicly available on a GitHub repository at \url{https://github.com/saref/clusterability-index} upon publication of the paper. 
	The R code and the processed data for analyzing and visualizing the results are publicly available on an OSF repository at \url{https://doi.org/10.17605/OSF.IO/3QTFB}.
	
	\subsection*{Solving the graph optimization models}
	
	The proposed optimization models can be solved by mathematical programming solvers which supports 0/1 linear programming (binary linear) models. The code for both optimization models will be made available on a GitHub repository at \url{https://github.com/saref/clusterability-index} 
	once this paper is published. In the GitHub repository, we provide Python code for using Gurobi solver (version 9.1) to solve the proposed binary linear models and obtain optimal partitions of signed networks into internally cohesive and mutually divisive clusters based on generalized balance theory. 
	
	\subsection*{An illustrative numerical example for the $k$-partitioning model}
	We provide a numerical example to illustrate how the mathematical programming model in Eq.\ 1 (in the paper) works (and how it is solved by a branch and bound algorithm). Consider that the model in  Eq.\ 1 is given the example signed graph of Fig.\ 1 (in the paper) and a the pre-defined value of $k=3$ for the number of clusters.

	The main role of the solver that solves this model is to explore the space of feasible solutions (feasible ways of clustering the input signed graph into $k$ clusters) and finding a feasible solution which is associated with the minimum number of frustrated edges. Without loss of generality, we can consider one step of this optimization process is evaluating the objective function value (the frustration count) for a given feasible solution. The following numerical example explains how the solver handles the model to complete this step and move forward if needed.

	Consider that the optimization solver is to evaluate the frustration count of the partition illustrated in  Fig.\ 1 (B). The non-zero $x_{ic}$ binary decision variables for this partition are as follows: $x_{1,1}=1$, $x_{2,1}=1$, $x_{3,1}=1$, $x_{4,2}=1$, and $x_{5,2}=1$. Every other $x_{ic}$ variable has to be $0$ for these variables to constitute a feasible solution (due to the first set of constraints of the model $\sum_{c \in C} x_{ic} = 1 \forall i \in V$).

	The second and third sets of constraints allow the model to determine the frustration status of each edge by quantifying all $f_{ij}$ variables based on the values of the $x_{ic}$ variables for the feasible solution under evaluation.

	For the positive edges $(1,3)$ and $(2,3)$, the second set of constraints ($f_{ij}  \ge  x_{ic} - x_{jc} ~\forall (i,j) \in E^+,  ~\forall c \in C$) is in place. These constraints for the feasible solution in Fig.\ 1 (B) lead to $f_{i,j}\geq 0$. Given the flexibility for taking either binary value, the minimization pressure from the objective function sets the values for $f_{1,3}$ and $f_{2,3}$ to $0$. This means that the edges $(1,3)$ and $(2,3)$ are not frustrated because they are positive and have the same cluster membership on their endpoints.

	For the negative edges $(1,4)$, $(1,5)$, $(2,5)$, and $(3,4)$, the third set of constraints ($f_{ij}  \ge  x_{ic} + x_{jc} -1 ~\forall (i,j) \in E^-,  ~\forall c \in C$) is in place. These constraints for the feasible solution in Fig.\ 1 (B) lead to $f_{i,j}\geq 0$. Given the flexibility for taking either binary value, the minimization pressure from the objective function sets the values for $f_{1,4}$, $f_{1,5}$, $f_{2,5}$, and $f_{3,4}$ to $0$. This means that the edges $(1,4)$, $(1,5)$, $(2,5)$, and $(3,4)$ are not frustrated because they are negative and have different cluster memberships on their endpoints.

	For the edge $(4,5)$, the third set of constraints is in place because it is a negative edge. The constraint associated with $c=2$ leads to $f_{i,j}\geq 1$ for the feasible solution in Fig.\ 1 (B). Therefore, $f_{4,5}$ takes the value $1$. This means that the edge $(4,5)$ is frustrated because it is negative and has the same cluster membership on its endpoints.

	Accordingly, the objective function $\sum_{(i,j) \in E}  f_{ij}$ is evaluated by the model to $1$ for the partition illustrated in Fig.\ 1 (B). As the linear programming relaxation of the model in Eq.\ 1 has a solution of $0$ for the signed graph in Fig.\ 1, the solver does not stop at this feasible solution and continues exploring other feasible solutions.

	At some point, it finds the feasible solution for the partition illustrated in Fig.\ 1 (C). The constraints of the model and the pressure from the minimization objective function lead to all $f_{i,j}$ variables taking the value $0$. Therefore, the objective function evaluates to $0$.

	At this stage of the branch and bound process, the upper bound (objective function of the best feasible solution found so far) and the lower bound (LP relaxation solution) reach each other and the solver stops and reports the partition illustrated in Fig.\ 1 (C) as an optimal $k$-partition for the input signed graph and the pre-defined parameter $k=3$.

	\subsection*{An illustrative numerical example for the partitioning model}
	We provide a numerical example to illustrate how the mathematical programming model in Eq.\ 2 (in the paper) works (and how it is solved by a branch and bound algorithm). Consider that the model in  Eq.\ 2 is given the example signed graph of Fig.\ 1 (in the paper).

	The main role of the solver that solves this model is to explore the space of feasible solutions (feasible ways of clustering the input signed graph into any number of clusters) and finding a feasible solution which is associated with the minimum number of frustrated edges. Without loss of generality, we can consider one step of this optimization process is evaluating the objective function value (the frustration count) for a given feasible solution. The following numerical example explains how the solver handles the model to complete this step and move forward if needed.

	Consider that the optimization solver is to evaluate the frustration count of the partition illustrated in  Fig.\ 1 (B). The non-zero $y_{ij}$ binary decision variables for this partition are as follows: $y_{1,2}=1$, $y_{1,3}=1$, $y_{2,3}=1$, and $y_{4,5}=1$. Every other $y_{ij}$ is $0$ because no other pairs of nodes are in the same cluster.

	Note that the term in the objective function for a positive edge is $1-y_{ij}$ because a positive edge is frustrated when its endpoints are in different clusters. The term in the objective function for a negative edge is $y_{ij}$ because a negative is frustrated when its endpoints are in the same cluster.

	Given the values of $y_{1,3}=1$ and $y_{2,3}=1$, the contribution of positive edges $(1,3)$ and $(2,3)$ to the objective function is $0$. This means that the positive edges $(1,3)$ and $(2,3)$ are not frustrated because they have the same cluster membership on their endpoints in the partition illustrated in  Fig.\ 1 (B).

	Given the value of $y_{4,5}=1$, the negative edge $(4,5)$ contributes $1$ to the objective function. This means that the negative edge $(4,5)$ is frustrated because it has the same cluster membership on its endpoints. The contribution of all other negative edges is $0$ because they all have different cluster memberships on their endpoints in the partition illustrated in  Fig.\ 1 (B).

	Accordingly, the objective function $\sum_{(i,j) \in E} a_{ij}((a_{ij}+1)/2) - a_{ij}y_{ij}$ is evaluated by the model to $1$ for the partition illustrated in Fig.\ 1 (B). As the linear programming relaxation of the model in Eq.\ 2 has a solution of $0$ for the signed graph in Fig.\ 2, the solver does not stop at this feasible solution and continues exploring other feasible solutions.

	At some point, it finds the feasible solution for the partition illustrated in Fig.\ 1 (C). The non-zero $y_{ij}$ binary decision variables for this partition are as follows: $y_{1,2}=1$, $y_{1,3}=1$, and $y_{2,3}=1$. Every other $y_{ij}$ is $0$ because no other pairs of nodes are in the same cluster. The objective function evaluates to $0$ because all positive edges have the same cluster membership on their endpoints and all negative edges have different cluster memberships on their endpoints.

	At this stage of the branch and bound process, the upper bound (objective function of the best feasible solution found so far) and the lower bound (LP relaxation solution) reach each other and the solver stops and reports the partition illustrated in Fig.\ 1 (C) as an optimal partition for the input signed graph.
	
	\subsection*{Using Gurobi for solving the proposed optimization models}
	
	Our proposed algorithms are developed in Python 3.8 based on the mathematical programming models discussed in the paper which partition signed networks based on generalized balance into an optimal $k$-partition or an optimal partition without specifying $k$.
	
	These optimization algorithms are distributed under an Attribution-NonCommercial-ShareAlike 4.0 International (CC BY-NC-SA 4.0) license. This means that one can use these algorithms for non-commercial purposes provided that they provide proper attribution for them by citing the current article. Copies or adaptations of the algorithms should be released under the similar license.
	
	The following steps outline the process for academics to install the required software (\textit{Gurobi} solver \cite{gurobi}) on their computer to be able to solve the optimization models:
	
	\begin{enumerate}
		\item 
		Download and install Anaconda (Python 3.8 version) which allows you to run a Jupyter code. It can be downloaded from \url{https://anaconda.com/products/individual}. Note that you must select your operating system first and download the corresponding installer.
		
		\item 
		Register for an account on \url{https://pages.gurobi.com/registration} to get a free academic license for using Gurobi. Note that Gurobi is a commercial software, but it can be registered with a free academic license if the user is affiliated with a recognized degree-granting academic institution. This involves creating an account on Gurobi website to be able to request a free academic license in step 5.
		
		\item 
		Download and install Gurobi Optimizer (versions 9.1 and above are recommended) which can be downloaded from \url{https://www.gurobi.com/downloads/gurobi-optimizer-eula/} after reading and agreeing to Gurobi's End User License Agreement.
		
		\item
		Install Gurobi into Anaconda. You do this by first adding the Gurobi channel to your Anaconda channels and then installing the Gurobi package from this channel.
		
		From a terminal window issue the following command to add the Gurobi channel to your default search list
		
		\begin{verbatim}
			conda config --add channels http://conda.anaconda.org/gurobi
		\end{verbatim} 
		
		Now issue the following command to install the Gurobi package
		
		\begin{verbatim}
			conda install gurobi
		\end{verbatim}
		
		\item 
		Request an academic license from \\ \url{http://gurobi.com/downloads/end-user-license-agreement-academic/} and install the license on your computer following the instructions given on Gurobi license page.
	\end{enumerate}

	Completing these steps is explained in the following links (for version 9.1).
	
	For Windows:
	
	\url{gurobi.com/documentation/9.1/quickstart_windows/index.html} 
	
	For Linux:
	
	\url{gurobi.com/documentation/9.1/quickstart_linux/index.html} 
	
	For Mac OS:
	
	\url{gurobi.com/documentation/9.1/quickstart_mac/index.html}.

	After following the instructions above, open Jupyter Notebook which takes you to an environment (a new tab on your browser pops up on your screen) where you can open the main code in a Jupyter notebook (which is a file with .ipynb extension).

\begin{figure}
		\centering
		\includegraphics[width=\textwidth]{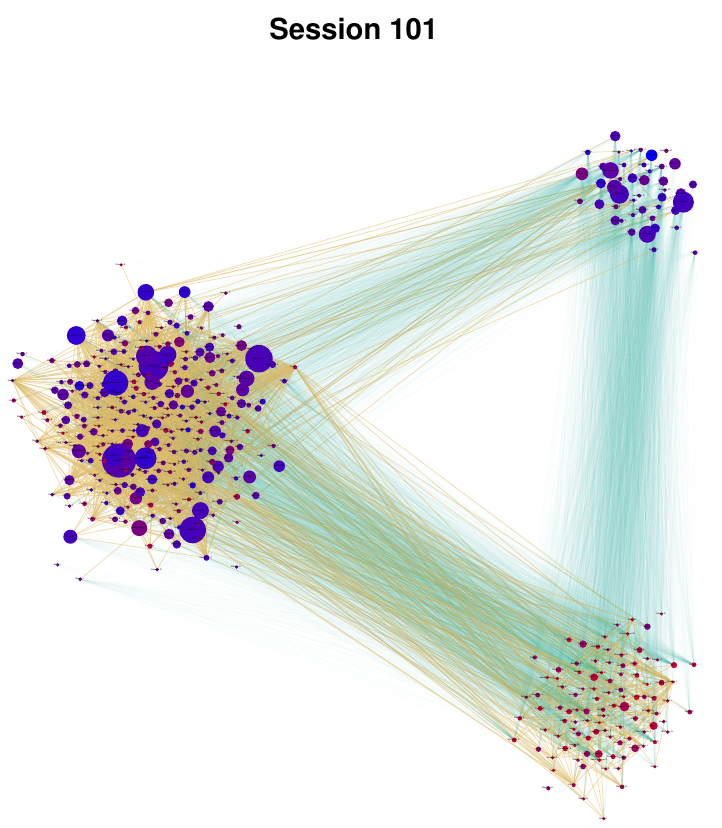}
		\caption{The 3-partition coalitions in the 101\textsuperscript{st} session of the House of Representatives}
		\label{H101}
	\end{figure}

\begin{figure}
		\centering
		\includegraphics[width=\textwidth]{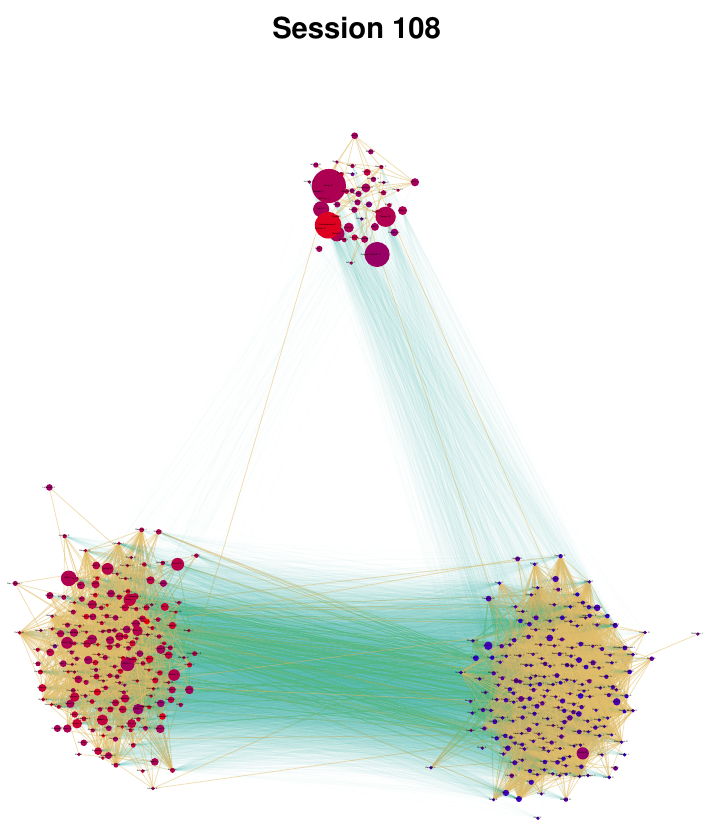}
		\caption{The 3-partition coalitions in the 108\textsuperscript{th} session of the House of Representatives}
		\label{H108}
	\end{figure}

	\subsection*{Visualization of 3-partition coalitions in selected House networks (Figures \ref{H101} and \ref{H108})}
	Fig. \ref{H101} shows the 3-partition coalitions in the 101\textsuperscript{st}, when the third coalition was dominated by highly effective ideologically liberal legislators. Fig. \ref{H108} shows the 3-partition coalitions in the 108\textsuperscript{th}, when the third coalition was dominated by highly effective ideologically conservative legislators. In both cases, brown (positive) and turquoise (negative) edges represent significantly many and significantly few co-sponsorships respectively. Node color indicates the legislator's ideology on a blue (liberal, Nokken-Poole = -1), purple (moderate, 0), red (conservative, +1) spectrum. Node size indicates the legislator's effectiveness. All nodes are labeled with legislators' names, which are visible when the figure is viewed at $400^+\%$ magnification.
	
	\clearpage
	
	\clearpage
	
	\clearpage

	\subsection*{Significance, generalizability, and limitations of our computational methods}
	The computational results we provided have broad relevance because they demonstrate the practical feasibility of solving fundamental NP-hard signed graph partitioning problems. Solving these partitioning problems are essential for exact evaluations of the structure of signed networks which go beyond the political science application we have demonstrated and have use cases in other fields from biology and physics \cite{iacono_determining_2010,huffner_separator-based_2010,facchetti2011,facchetti2012exploring,tahmassebi2019determining,li2019binary} to social sciences \cite{harary_signed_2002,patrick_doreian_structural_2015,souto2018capturing,derr2018congressional,neal2018,aref2020detecting,arinik2020multiple,aref2020multilevel,schoch2020legislators}. Specifically, our methods for partitioning a signed graph according to generalized balance improve upon heuristic methods that are fast but do not generally yield optimal partitions \cite{drummond2013efficient,levorato2015ils,levorato2017evaluating}. Additionally, our methods also improve upon existing methods for obtaining optimal partitions that are only capable of handling small graphs with $n \leq 40$ \cite{brusco_k-balance_2010,figueiredo2013mixed}. The correctness of our methods for partitioning signed networks is guaranteed by the branch and bound algorithm of Gurobi \cite{gurobi} which is an exact method for solving binary linear programming models to global optimality.
	
	The sizes of real-world instances we have solved to global optimality are considerable and therefore suggest that our proposed models can be used for a wide variety of other applications with networks of similar and smaller sizes. For example, the network of the 115\textsuperscript{th} session has $n=448$ nodes, $m=31,936$ edges, and $|T|=9,134,395$ node triples with at least one edge. Obtaining an optimal $7$-partition using Eq.\ 1 leads to an optimization model with $nk+m=35,072$ binary variables and $mk+n=224,000$ constraints, which takes Gurobi, 1.66 hours\footnote{23.58 minutes when solving this instance again using Gurobi 9.5.2 on a laptop with 11th Gen Intel(R) Core(TM) i7-11800H @ 2.30GHz and 64 GBs of RAM running Windows 10 Home} to solve. Moreover, obtaining an optimal partition (without specifying $k$) using Eq.\ 2 leads to an optimization model with $n(n-1)/2=100,128$ binary variables and $3|T|=27,403,185$ constraints, which takes Gurobi only 5.28 hours\footnote{only 10.28 minutes when solving this instance again using Gurobi 9.5.2 on a laptop with 11th Gen Intel(R) Core(TM) i7-11800H @ 2.30GHz and 64 GBs of RAM running Windows 10 Home} to solve. While obtaining these partitions requires a few hours\footnote{a few minutes; using Gurobi 9.5.2 on a laptop with 11th Gen Intel(R) Core(TM) i7-11800H @ 2.30GHz and 64 GBs of RAM running Windows 10 Home}, the resulting partition is guaranteed to be globally optimal, which is essential for an exact evaluation of the structure of the signed networks under analysis. 
	
	As expected from the NP-hardness of the problems, the main limitation of the models in Eqs.\ 1--2 (in the manuscript) is the size of the network they can handle in a reasonable time. We have demonstrated the practicability of these models for real-world political networks with up to $\sim 30,000$ edges considering that a few hours\footnote{a few minutes; using Gurobi 9.5.2 on a laptop with 11th Gen Intel(R) Core(TM) i7-11800H @ 2.30GHz and 64 GBs of RAM running Windows 10 Home} is worth finding a globally optimal solution for the exact evaluation of the structure of these network. From a practical standpoint, two factors are relevant for determining whether these computationally intensive models are suitable for a different use case: network properties and processing capabilities. Previous studies suggest that some properties of the input graph like degree heterogeneity could be determinant factors of solve time in similar problems \cite{aref2016exact}. Also, structural regularities in networks constructed from empirical data often make them easier to solve compared to synthetic networks (like random graphs) \cite{aref2016exact}. As Gurobi solver makes use of multiple processing threads to explore the feasible space in parallel, the processing capabilities of the computer that runs the optimization solver could also make an impact. Therefore, our experiments do not guarantee that every network with up to $30,000$ edges can be optimally partitioned based on generalized balance within the solve times that we have observed for our real-world instances of US House signed networks.
	
	The computing processor configuration we have used (32 Intel Xeon CPU E7-8890 v3 @ 2.50 GHz processors) and the size of the real networks we have analyzed ($m \sim 30,000$) have led to solve times of roughly a few hours\footnote{a few minutes; using Gurobi 9.5.2 on a laptop with 11th Gen Intel(R) Core(TM) i7-11800H @ 2.30GHz and 64 GBs of RAM running Windows 10 Home} per instance. One could speculate that larger networks on the same hardware or the same networks on less powerful hardware is expected to take longer. In such cases, one may consider using a non-zero optimality gap tolerance (\textit{MIPGap} as a Gurobi parameter \cite{gurobi}) to find solutions within a guaranteed proximity of optimality to reduce the solve time.

	\subsection*{Multiplicity of optimal solutions}
	
	There are symmetries in the mathematical formulations for the two models in Eqs.\ 1--2 (in the manuscript). For example, in Eq.\ 1, a given 2-partition can be expressed by different feasible solutions (sets of values for decision variables). This is because the clusters are treated indifferently and could be swapped while the partition remains virtually unchanged. As another example, in Eq.\ 2, a feasible solution does not necessarily represent a unique partition. This is due to the original formulation \cite{demaine2006correlation} in which a pair of non-adjacent nodes $a$ and $i$ may have no decision variables indicating they belong to the same cluster with any of their neighbours (denoted by $b$ and $j$ respectively $\forall b,j: a_{ab}\neq0 , a_{ij}\neq0$), i.e., all the decision variables associated with $a$ and $i$ take the value zero. In that case, the same feasible solution could lead to two partitions (with identical fitness) depending on whether nodes $a$ and $i$ are placed in the same or different clusters. Another source of symmetry is the existence of isolate nodes whose optimal cluster membership is random and therefore not meaningful. When characterizing the composition of clusters in our analyses, we have ignored isolates.
	
	Due to the symmetries outlined above, both optimization models in Eqs.\ 1--2 generally have multiplicity in their optimal solutions. Finding all optimal solutions to such computationally intensive problems are not practically feasible for large instances. For small instances, however, previous studies have looked at multiplicity of optimal solutions in similar partitioning problems \cite{aref2020multilevel,arinik2021multiplicity}. Although optimal 2-partitions can be unique in some small real-world signed networks \cite{aref2020multilevel}, more often multiple optimal solutions exist \cite[their Fig. S1]{aref2020multilevel}. Also, in the case of small complete random signed graphs, multiple optimal solutions may exist \cite{arinik2021multiplicity}. Due to the practical complexity of these problems and the size of empirical networks we consider, although we cannot find and analyze all optimal partitions, it is certain that the optimal partitions are not unique. Future work is needed to find practical methods for finding and analyzing all optimal partitions of such large networks.

	\subsection*{Oppositional ties of the splinter coalition}
	Members of the third coalition have 21.18 negative ties for every positive tie which is substantially different from the members of traditional coalitions who have on average 2.68 negative ties for every positive tie. This distinction in oppositional ties deserves more attention and we look at the fraction of each type of edge by coalition, taking into account the party of legislator at the other endpoint of the edge.
	
\begin{figure}[htb!]
		\centering
		\includegraphics[width=1\textwidth]{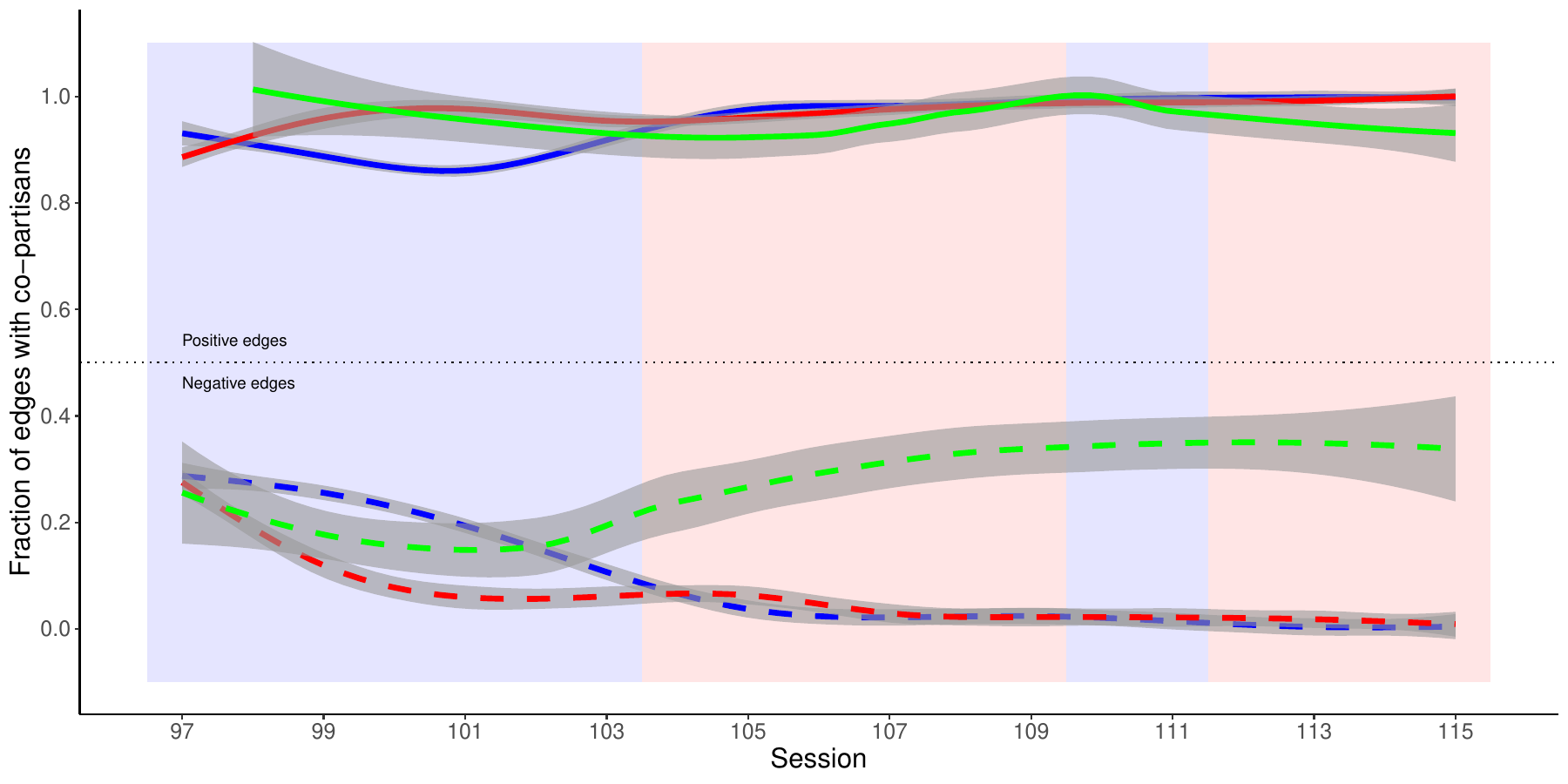}
		\caption{Fractions of positive and negative edges to members of the same party aggregated for each of the three coalitions}
		\label{fig:negative}
	\end{figure}

	Figure \ref{fig:negative} illustrates the fractions of positive and negative edges with co-partisans (members of the same party) for each of the coalitions based on the optimal 3-partitions. Fractions of positive (negative) edges are shown by solid (dashed) lines. The red, blue, and green lines represent the conservative coalition, the liberal coalition, and the splinter coalition respectively. It can be seen in Figure \ref{fig:negative} that the three coalitions are similar based on the fraction of positive edges with co-partisans: members of all coalitions mainly collaborate (i.e. have a positive edge with) members of their own party. For the fraction of negative edges with co-partisans, however, the splinter coalition shifts away from the main liberal and conservative coalition. From the 104\textsuperscript{th} session, this quantity has generally increased for the splinter coalition reaching values close to $0.4$. This means that legislators in the splinter coalitions have a considerable proportion (nearly $40\%$) of their negative edges with members of the own party. Given this distinctive feature in oppositional ties, one may conclude that the members of the third coalition are distinctively more willing to push back against their own party. 
	
	%\clearpage
	
	\FloatBarrier

	\subsection*{Additional numerical results (Tables \ref{tab:housenetworks} to \ref{tab:housetable})}

	\begin{sidewaystable}\centering
		\caption{Detailed properties and clusterability indices for networks}
		\resizebox{\textwidth}{!}{%
			\begin{tabular}{lllllllllllllll}
				\hline
				Session & Start year & $n$ & $m$   & density & $m^-$ & $m^+$ & $m^-/m$ & $C_2(G)$ & $C_3(G)$   & $C_4(G)$     & $C_5(G)$     & $C_6(G)$     & $C_7(G)$     & $C(G)$       \\ \hline
				97      & 1981 & 447 & 4954  & 0.05    & 4494  & 460   & 0.91    & 280      & 36         & 14           & 6            & \textbf{4}   & \textbf{4}   & \textbf{4}   \\
				98      & 1983 & 444 & 6744  & 0.07    & 5460  & 1284  & 0.81    & 435      & 102        & 60           & 49           & 45           & \textbf{44}  & \textbf{44}  \\
				99      & 1985 & 443 & 8714  & 0.07    & 6759  & 1955  & 0.78    & 962      & 79         & 48           & 40           & \textbf{38}  & \textbf{38}  & \textbf{38}  \\
				100     & 1987 & 446 & 8520  & 0.08    & 6888  & 1632  & 0.81    & 560      & 118        & 89           & \textbf{84}  & \textbf{84}  & \textbf{84}  & \textbf{84}  \\
				101     & 1989 & 449 & 10180 & 0.10    & 7970  & 2210  & 0.78    & 896      & 263        & 207          & 188          & \textbf{183} & \textbf{183} & \textbf{183} \\
				102     & 1991 & 447 & 13692 & 0.11    & 10152 & 3540  & 0.74    & 748      & 169        & 119          & 112          & \textbf{111} & \textbf{111} & \textbf{111} \\
				103     & 1993 & 446 & 13281 & 0.10    & 10530 & 2751  & 0.79    & 301      & 245        & \textbf{241} & \textbf{241} & \textbf{241} & \textbf{241} & \textbf{241} \\
				104     & 1995 & 445 & 13019 & 0.13    & 10301 & 2718  & 0.79    & 17       & \textbf{7} & \textbf{7}   & \textbf{7}   & \textbf{7}   & \textbf{7}   & \textbf{7}   \\
				105     & 1997 & 449 & 16709 & 0.13    & 13125 & 3584  & 0.79    & 50       & 22         & \textbf{16}  & \textbf{16}  & \textbf{16}  & \textbf{16}  & \textbf{16}  \\
				106     & 1999 & 442 & 17746 & 0.14    & 13390 & 4356  & 0.75    & 91       & 48         & 40           & \textbf{39}  & \textbf{39}  & \textbf{39}  & \textbf{39}  \\
				107     & 2001 & 447 & 17908 & 0.14    & 14062 & 3846  & 0.79    & 59       & 34         & \textbf{31}  & \textbf{31}  & \textbf{31}  & \textbf{31}  & \textbf{31}  \\
				108     & 2003 & 444 & 19384 & 0.16    & 14211 & 5173  & 0.73    & 63       & 33         & \textbf{29}  & \textbf{29}  & \textbf{29}  & \textbf{29}  & \textbf{29}  \\
				109     & 2005 & 445 & 21601 & 0.17    & 15684 & 5917  & 0.73    & 69       & 44         & 35           & \textbf{34}  & \textbf{34}  & \textbf{34}  & \textbf{34}  \\
				110     & 2007 & 452 & 23719 & 0.15    & 17094 & 6625  & 0.72    & 44       & 31         & \textbf{30}  & \textbf{30}  & \textbf{30}  & \textbf{30}  & \textbf{30}  \\
				111     & 2009 & 451 & 21564 & 0.21    & 15342 & 6222  & 0.71    & 153      & 87         & 82           & \textbf{78}  & \textbf{78}  & \textbf{78}  & \textbf{78}  \\
				112     & 2011 & 450 & 29904 & 0.19    & 21118 & 8786  & 0.71    & 39       & 30         & \textbf{28}  & \textbf{28}  & \textbf{28}  & \textbf{28}  & \textbf{28}  \\
				113     & 2013 & 447 & 25992 & 0.21    & 19026 & 6966  & 0.73    & 24       & 20         & \textbf{18}  & \textbf{18}  & \textbf{18}  & \textbf{18}  & \textbf{18}  \\
				114     & 2015 & 446 & 30149 & 0.22    & 21299 & 8850  & 0.71    & 57       & 38         & \textbf{37}  & \textbf{37}  & \textbf{37}  & \textbf{37}  & \textbf{37}  \\
				115     & 2017 & 448 & 31936 & 0.32    & 21883 & 10053 & 0.69    & 221      & 38         & \textbf{35}  & \textbf{35}  & \textbf{35}  & \textbf{35}  & \textbf{35}  \\ \hline
				\multicolumn{15}{l}{Cases indicating the stagnation of frustrated edges are shown in bold-face font.} \\
			\end{tabular}%
		}
		\label{tab:housenetworks}
	\end{sidewaystable}
	
	\begin{sidewaystable}\centering
		\caption{The size and effectiveness of the two parties and optimal 3-partition coalitions}
		\label{tab:housetable}
		\resizebox{\textwidth}{!}{%
			\begin{tabular}{l|cccccc||ccccccccc}
				\hline
				& \multicolumn{6}{c}{\textbf{Partition based on political party}} & \multicolumn{9}{c}{\textbf{Generalized Balance Partition ($k=3)$}} \\
				& \multicolumn{2}{c}{\underline{Size}} & \multicolumn{2}{c}{\underline{Median Ideology}} & \multicolumn{2}{c}{\underline{Mean Effectiveness}} & \multicolumn{3}{c}{\underline{Size}} & \multicolumn{3}{c}{\underline{Median Ideology}} & \multicolumn{3}{c}{\underline{Mean Effectiveness}} \\
				session & D & R & D & R & D & R & L & C & S & L & C & S & L & C & S \\
				\hline
				97 & 245 & 194 & -0.33 & 0.28 & 1.45 & 0.44 & 153 & 255 & 18 & -0.40 & 0.20 & 0.29 & 1.29 & 0.83 & 1.59 \\
				98 & 272 & 167 & -0.32 & 0.32 & 1.37 & 0.39 & 288 & 99 & 51 & -0.22 & 0.32 & -0.37 & 1.05 & 0.55 & 1.55 \\
				99 & 256 & 182 & -0.32 & 0.34 & 1.45 & 0.37 & 304 & 85 & 47 & -0.19 & 0.41 & -0.41 & 1.07 & 0.40 & 1.70 \\
				100 & 261 & 179 & -0.33 & 0.35 & 1.40 & 0.43 & 282 & 125 & 29 & -0.28 & 0.39 & -0.34 & 1.09 & 0.54 & 2.31 \\
				101 & 264 & 179 & -0.33 & 0.35 & 1.41 & 0.40 & 290 & 105 & 45 & -0.22 & 0.42 & -0.37 & 1.10 & 0.40 & 1.70 \\
				102 & 270 & 170 & -0.33 & 0.34 & 1.40 & 0.39 & 299 & 100 & 37 & -0.24 & 0.41 & -0.36 & 1.11 & 0.42 & 1.94 \\
				103 & 259 & 180 & -0.34 & 0.38 & 1.45 & 0.32 & 224 & 176 & 32 & -0.36 & 0.38 & -0.21 & 1.53 & 0.35 & 0.89 \\
				104 & 207 & 231 & -0.38 & 0.38 & 0.33 & 1.60 & 140 & 258 & 22 & -0.44 & 0.36 & -0.31 & 0.31 & 1.39 & 0.49 \\
				105 & 211 & 231 & -0.39 & 0.39 & 0.33 & 1.62 & 194 & 208 & 32 & -0.40 & 0.41 & 0.15 & 0.40 & 1.50 & 1.23 \\
				106 & 212 & 224 & -0.39 & 0.39 & 0.39 & 1.58 & 210 & 182 & 42 & -0.39 & 0.43 & 0.23 & 0.41 & 1.41 & 2.21 \\
				107 & 213 & 228 & -0.39 & 0.38 & 0.41 & 1.54 & 204 & 202 & 30 & -0.39 & 0.40 & 0.24 & 0.44 & 1.42 & 1.96 \\
				108 & 208 & 230 & -0.39 & 0.41 & 0.40 & 1.53 & 205 & 186 & 46 & -0.39 & 0.43 & 0.29 & 0.43 & 1.27 & 2.38 \\
				109 & 203 & 236 & -0.38 & 0.41 & 0.48 & 1.51 & 210 & 210 & 15 & -0.38 & 0.42 & 0.25 & 0.48 & 1.46 & 1.63 \\
				110 & 241 & 206 & -0.38 & 0.43 & 1.45 & 0.45 & 242 & 195 & 8 & -0.38 & 0.44 & 0.12 & 1.43 & 0.44 & 0.69 \\
				111 & 263 & 182 & -0.37 & 0.47 & 1.36 & 0.44 & 195 & 176 & 69 & -0.42 & 0.47 & -0.18 & 1.34 & 0.44 & 1.42 \\
				112 & 198 & 245 & -0.40 & 0.47 & 0.47 & 1.42 & 178 & 250 & 9 & -0.41 & 0.46 & 0.29 & 0.47 & 1.33 & 1.75 \\
				113 & 203 & 238 & -0.40 & 0.48 & 0.51 & 1.42 & 197 & 236 & 4 & -0.41 & 0.48 & 0.06 & 0.52 & 1.43 & 0.34 \\
				114 & 188 & 249 & -0.40 & 0.49 & 0.58 & 1.32 & 186 & 242 & 8 & -0.40 & 0.50 & 0.21 & 0.58 & 1.28 & 2.20 \\
				115 & 195 & 243 & -0.39 & 0.49 & 0.59 & 1.35 & 192 & 211 & 33 & -0.39 & 0.52 & 0.24 & 0.59 & 1.27 & 1.80 \\
				\hline
				\multicolumn{16}{l}{In party-partition: D = Democratic coalition, R = Republican coalition} \\
				\multicolumn{16}{l}{In 3-partition: L = Liberal coalition, C = Conservative coalition, S = Splinter coalition}
			\end{tabular}%
		}
	\end{sidewaystable}

	%%% Add this line AFTER all your figures and tables
	\FloatBarrier
	
	\subsection*{Movie: Slideshow of the 3-partition coalitions of the signed US House networks}
	A slideshow of optimal 3-partition coalitions is available online at \url{https://saref.github.io/SI/AN2021/House_coalitions.mp4}
	which includes all 19 House networks. Green and red edges represent significantly many and significantly few co-sponsorships respectively. Node color indicates the legislator's ideology on a blue (liberal, Nokken-Poole = -1), purple (moderate, 0), red (conservative, +1) spectrum. Node size indicates the legislator's effectiveness. Looking at the colors and positions of edges we can see that the large majority of edges are intra-cluster positive or inter-cluster negative. In these networks, only $0.05\%$--$2.5\%$ of the edges are frustrated under the optimal 3-partitions which indicate the closeness of the networks to the assertions of the generalized balance theory \cite{davis1967clustering}. If we look at the colors of the nodes, we see the ideological divide between the members of different coalitions. The splinter coalition is the smallest cluster of the nodes which usually has several large nodes (highly effective legislators).\\
	
	\subsection*{Dataset: frustrated\_legislators.RData and frustrated\_legislators.R on OSF}
	The file `frustrated\_legislators.RData' is an R workspace which includes a dataframe object `data' that contains details about each legislator in each session (e.g. ideology, effectiveness, cluster membership in optimal $k$-partitions), and 19 igraph objects `H\#\#\#' that contain signed networks for each session. The file `frustrated\_legislators.R' in the same repository contains the R code to replicate all substantive analyses reported in the manuscript using these data. Both files are publicly available at \url{https://doi.org/10.17605/OSF.IO/3QTFB}. The data are distributed under a CC-BY 4.0 license, which means that they can be used provided they are properly attributed by citing \cite{neal2018,domagalski2021} and the current article.

	\subsection*{Dataset: clusters-house.csv}
	The results on globally optimal solutions to the optimization model for $k$-partitioning House networks are available in comma-separated values format at \url{saref.github.io/SI/AN2021/clusters-house.csv}.
	The first and second columns contain session numbers and legislator name as indicated by the headers. Each row is a legislator-session combination. The other columns are the cluster assignments based on optimal $k$-partitions for $k \in \{2,3,\dots,7\}$ as indicated by the column header. The entries represent the cluster assignment of the node associated to the row (the legislator-session combination) based on an optimal solution of the $k$-partition associated to the column.

\end{document}